%% file: 2256_2018.tex
\documentclass[fleqn,usenatbib]{mnras}

\usepackage{newtxtext,newtxmath}

\usepackage[T1]{fontenc}
\usepackage{ae,aecompl}

\usepackage{graphicx}	
\usepackage{amsmath}	
\usepackage{amssymb}	
\usepackage{nicefrac}	
\graphicspath{ {./images/} }   
\usepackage{pdflscape}
\usepackage{subfig}
\usepackage{siunitx}
\usepackage{threeparttable}
\usepackage{gensymb}
\usepackage{hyperref}

\usepackage{tabularx} 
\usepackage{multirow}

\input{macros}

\title[The magnetic field of the eclipsing material of PSR~J2256--1024]{The GBT 350-MHz Drift Scan Pulsar Survey. III. Detection of a magnetic field in the eclipsing material of PSR~J2256--1024}

\author[K. Crowter et al.]{Kathryn Crowter$^{1}$\thanks{E-mail: kathryn.crowter@gmail.com},
Ingrid H. Stairs$^{1}$,
Christie A. McPhee$^{1}$,
\newauthor
Anne M. Archibald$^{2,3}$,
Jason Boyles$^{4}$,
Jason Hessels$^{2,3}$,
Chen Karako-Argaman$^{5}$,
\newauthor
Duncan R. Lorimer$^{6,7}$,
Ryan S. Lynch$^{8,7}$,
Maura A. McLaughlin$^{6,7}$,
\newauthor
Scott M. Ransom$^{9}$,
Mallory S.E. Roberts$^{10,11}$,
Kevin Stovall$^{12}$,
\newauthor
and Joeri van Leeuwen$^{3,2}$
\\
$^{1}$Department of Physics and Astronomy, University of British Columbia, 6224 Agricultural Road, Vancouver, BC, V6T 1Z1, Canada\\
$^{2}$Anton Pannekoek Institute for Astronomy, University of Amsterdam, Science Park 904, 1098 XH Amsterdam, The Netherlands\\
$^{3}$ASTRON, Netherlands Institute for Radio Astronomy, Oude Hoogeveensedijk 4, 7991 PD Dwingeloo, The Netherlands\\
$^{4}$Department of Physics and Astronomy, Western Kentucky University, 1906 College Heights Blvd., Bowling Green, \\KY, 42101, USA\\
$^{5}$Department of Physics and McGill Space Institute, McGill University, Montreal, QC, Canada H3A 2T8\\
$^{6}$Department of Physics and Astronomy, West Virginia University, Morgantown, WV 26506, USA\\
$^{7}$Center for Gravitational Waves and Cosmology, West Virginia University, Chestnut Ridge Research Building, Morgantown, \\WV 26505, USA\\
$^{8}$Green Bank Observatory, PO Box 2, Green Bank, WV 24944, USA\\
$^{9}$National Radio Astronomy Observatory, Charlottesville, VA 22903, USA\\
$^{10}$New York University Abu Dhabi, Saadiyat Island, Abu Dhabi, UAE\\
$^{11}$Eureka Scientific, Inc. Oakland, CA USA\\
$^{12}$National Radio Astronomy Observatory, 1003 Lopezville Road, Socorro, NM 87801, USA
}

\date{Accepted XXX. Received YYY; in original form ZZZ}

\pubyear{2020}

\begin{document}
\label{firstpage}
\pagerange{\pageref{firstpage}--\pageref{lastpage}}
\maketitle

\begin{abstract}
We present the first measurement of a non-zero magnetic field in the eclipsing material of a black widow pulsar. Black widows are millisecond pulsars which are ablating their companions; therefore they are often proposed as one potential source of isolated millisecond pulsars. PSR~J2256--1024 is an eclipsing black widow discovered at radio wavelengths and later also observed in the X-ray and gamma parts of the spectrum. Here we present the radio timing solution for PSR~J2256--1024, polarization profiles at 350, 820, and 1500~MHz and an investigation of changes in the polarization profile due to eclipsing material in the system. In the latter we find evidence of Faraday rotation in the linear polarization shortly after eclipse, measuring a rotation measure of \SI{0.44 \pm 0.06}{\rad\per\meter\squared} and a corresponding line-of-sight magnetic field of $\sim$\SI{3.5 \pm 1.7}{\milli\gauss}. \end{abstract}

\begin{keywords}
 pulsars: general --  pulsars: J2256--1024 -- magnetic field -- polarization
\end{keywords}


\section{Introduction}
\label{sec:intro}
Black widow pulsars are millisecond pulsars (MSPs) in short orbits (\begin{math}P_{\rm B}<1\text{d}\end{math}) with low mass companions (\begin{math}M_{\rm C} \ll \SI{0.1}{\msun}\end{math}), such that the companion is being gradually destroyed by the pulsar's wind. The first such system, PSR~B1957+20, was discovered in 1988~\citep{1988Natur.333..237F} showing asymmetrical, radio eclipses larger than the companion's Roche lobe. This prompted the idea of a classic black widow; material blown from the companion forms a cloud around the star which, geometry willing, then also blocks the pulsar signal. 

At first these pulsars seemed comparatively rare in the Galactic field (as opposed to globular clusters, where they are more common) as discoveries trickled in: PSR J2051--0827 in 1996~\citep{Stappers1996a} and PSR J0610--2100 in 2006~\citep{Burgay2006a}. The situation changed dramatically after the launch of the {\em Fermi} Gamma-ray Space Telescope in 2008; many bright gamma-ray sources were found with {\em Fermi}'s Large Area Telescope (LAT)~\citep{Atwood2009} that were subsequently identified as black widow pulsars. Now, black widows and the so-called ``redbacks'' (a similar group with higher mass companions) make up \begin{math} \approx \SI{10}{\percent} \end{math} of the current sample of 300 Galactic millisecond pulsars~\citep{Manchester_2005}\footnote{http://www.atnf.csiro.au/research/pulsar/psrcat}.

PSR J2256--1024 (hereafter J2256--1024) is a black widow MSP discovered in the Green Bank Observatory's Robert C. Byrd Green Bank Telescope (GBT) 350-MHz Drift-Scan Survey~\citep{Boyles:2011,Lynch2013}. After the initial radio detection, J2256--1024 was also detected as a gamma-ray source by {\em Fermi} LAT~\citep{Abdo2010} and has since been confirmed as a gamma-ray pulsar~\citep{Bangale2011,Abdo2013}. In the X-ray part of the spectrum, \citet{Gentile2013} found J2256--1024 to emit photons both from the surface of the neutron star with a blackbody spectrum (of flux $2.5_{-1.0}^{+1.0}~ 10^{-14}$\si{\erg\per\second\per\centi\meter\squared}), and from an intra-binary shock, with a power-law spectrum (of flux $3.3_{-1.8}^{+2.6}~ 10^{-14}$\si{\erg\per\second\per\centi\meter\squared} and index $1.8_{-0.6}^{+0.7}$). 

Additionally, an optical companion to J2256--1024 was discovered in the \citet{1302.1790} investigation into strongly irradiated companions of certain {\em Fermi}-detected MSPs. \citet{1302.1790} find an inclination angle for the orbit of \ang{68\pm11} and a size for the companion ``not inconsistent with a solar-composition, degenerate object".

\autoref{sec:obs_data} covers data acquisition and reduction. In \autoref{sec:timing} we present the timing solution for J2256--1024 based on 3 years of radio observations with the GBT, plus \autoref{subsec:gamma_stuff} describes a gamma-ray analysis using $\approx 10.8$ years of {\em Fermi} LAT photons. \autoref{sec:pol_profiles} presents polarization profiles at 350, 820, and~1500 MHz. Dynamic spectra at 350 and 820 MHz are presented in \autoref{sec:dyn_spec} along with some measured spectral properties. Finally in \autoref{sec:eclipse} we discuss changes to the polarization profiles near the eclipse and use these changes to measure a magnetic field within the eclipsing material. A preliminary version of this analysis was presented in \citet{Crowter2018}.

\section{Observations and Data Reduction}
\label{sec:obs_data}
Observations used in this analysis were all taken with the GBT. There are 31 epochs between modified Julian dates (MJDs) 55005 and 56093 (2009 June 23 --  2012 June 15) ranging over several GBT project numbers, with full polarimetric data on 16 epochs. Three dual-linear polarization receivers were used at the GBT: one located at the prime focus covering \SIrange[range-phrase = --]{290}{395}{\MHz}; another, also at the prime focus, operating over \SIrange[range-phrase = --]{680}{920}{\MHz}; and the Gregorian ``L-band'' receiver spanning \SIrange[range-phrase = --]{1150}{1750}{\MHz}.

The majority of the data were taken using the Green Bank Ultimate Pulsar Processing Instrument \citep[GUPPI;][]{2009AAS...21460508R} backend. However on three epochs (MJDs 55181, 55191, and 55226 at \SI{820}{\MHz}, \SI{350}{\MHz} and L-band respectively) observations covered the entire orbital phase of the system and data were recorded using both Green Bank Astronomical Signal Processor (GASP)~\citep{Demorest:2007} and GUPPI backends concurrently. All data taken using GASP were folded and dedispersed coherently whereas those taken with GUPPI underwent incoherent dedispersion. GUPPI has a much higher bandwidth than GASP; for example, at \SI{820}{\MHz} GUPPI has \SI{200}{\MHz} of bandwidth whereas GASP has \SI{64}{\MHz}. GUPPI data taken before MJD 54999 were excluded due to a known error in the field-programmable gate array (FPGA) code.


\begin{figure}
	\includegraphics[]{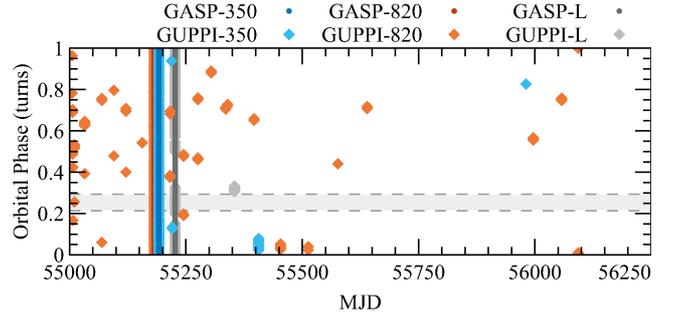}
    \caption{Data span and coverage. Labels denote the backend and central frequency of the observation. The grey band shows the approximate location of the eclipse, phases 0.215 - 0.293.}
    \label{fig:ophase_v_mjd}
\end{figure}


\autoref{fig:ophase_v_mjd} shows the coverage of the data in time, frequency, and orbital phase. As the primary aim of observations was to improve upon the initial timing solution for J2256--1024, observations other than the three epochs mentioned above were scheduled away from the eclipse region. Most of the observations were taken with a central frequency of \SI{820}{\MHz} and multi-frequency epochs are rare. There are also some notable gaps in MJD coverage over the data span.

Data were reduced using the \prog{PSRCHIVE} software suite \citep{2004PASA...21..302H}. The module \prog{paz} was used to zero-weight frequency channels at the band edges, where the signals are known to be depolarised due to quantization distortions, and to excise radio-frequency interference (RFI) in individual frequency channels and sub-integrations.

Before most observations a calibration (``cal'') observation was taken, using a gated noise diode to inject a known signal into the signal path. We performed a flux calibration (``fluxcal'') using \prog{fluxcal} from two cal files, one pointing at a strong source for which the flux is known, in this case QSO B1445+101, and one pointing a degree or two off that source. In this analysis observations with accompanying cal files were calibrated using \prog{pac -x}. This algorithm assumes the polarization feeds are perfectly orthogonal and combining this model with the fluxcal (if available) allows for differences in how each polarization feed is illuminated by the source.

\subsection{Polarization Profiles}
\label{subsec:data_reduction_pol_profiles}
Polarization profiles were produced from the three long-duration observation epochs at MJDs 55181, 55191, and 55226. \prog{rmfit}'s iterative algorithm was used on the GUPPI data at each frequency which provided rotation measure (RM) measurements. Data at each epoch were RM corrected based on the RM measurement made from the GUPPI data at that epoch. For each backend and frequency combination, all frequency channels were then summed together with \prog{pam}, the eclipse and surrounding regions were excised, summed into 512 bins across the pulse profile, and finally the remaining sub-integrations were summed together to form the total polarization profile.

\subsection{TOA generation}
\label{subsec:data_reduction_TOAs}
In several cases data taken using the same receiver and backend combination were taken with differing numbers of phase bins or frequency channels. As a result, data were binned to the lowest number within that backend-receiver subset. In the few cases where frequency channel binning was necessary, the dispersion measure (DM) was set to zero before binning and restored to its true value afterwards in order to correctly mimic incoherently dedispersed filterbank data with a smaller number of channels. Standard profiles for each backend and frequency combination were made by summing calibrated, eclipse-excised observations together. The profiles were then summed over frequency and sub-integration before being smoothed and aligned. Pulsar times-of-arrival (TOAs) were derived from profiles formed by binning the observations to one frequency channel, then binning in time such that the time per sub-integration was between approximately 0.3\% and 
2\% of the orbital period. The TOAs were measured by cross-correlating the total intensity profile in each sub-integration with that of its standard with \prog{pat} using the fast Fourier transform (FFT) algorithm~\citep{Taylor:1992}.

\subsection{Eclipse Analysis}
\label{subsec:data_reduction_eclipse}
J2256--1024's transition into and out of eclipse is fairly fast.
Our study of the eclipse and its surroundings was a balance between wanting a higher signal-to-noise ratio (S/N) and high time resolution. When timing J2256--1024, observations covering the eclipse had been binned in time by a factor of 16 to generate TOAs. Once a final timing solution was arrived at, higher-time-resolution TOAs were generated from the unbinned files for MJDs 55181, 55191, and 55226.

\section{Timing Solution}
\label{sec:timing}
\begin{figure*}
	\includegraphics[]{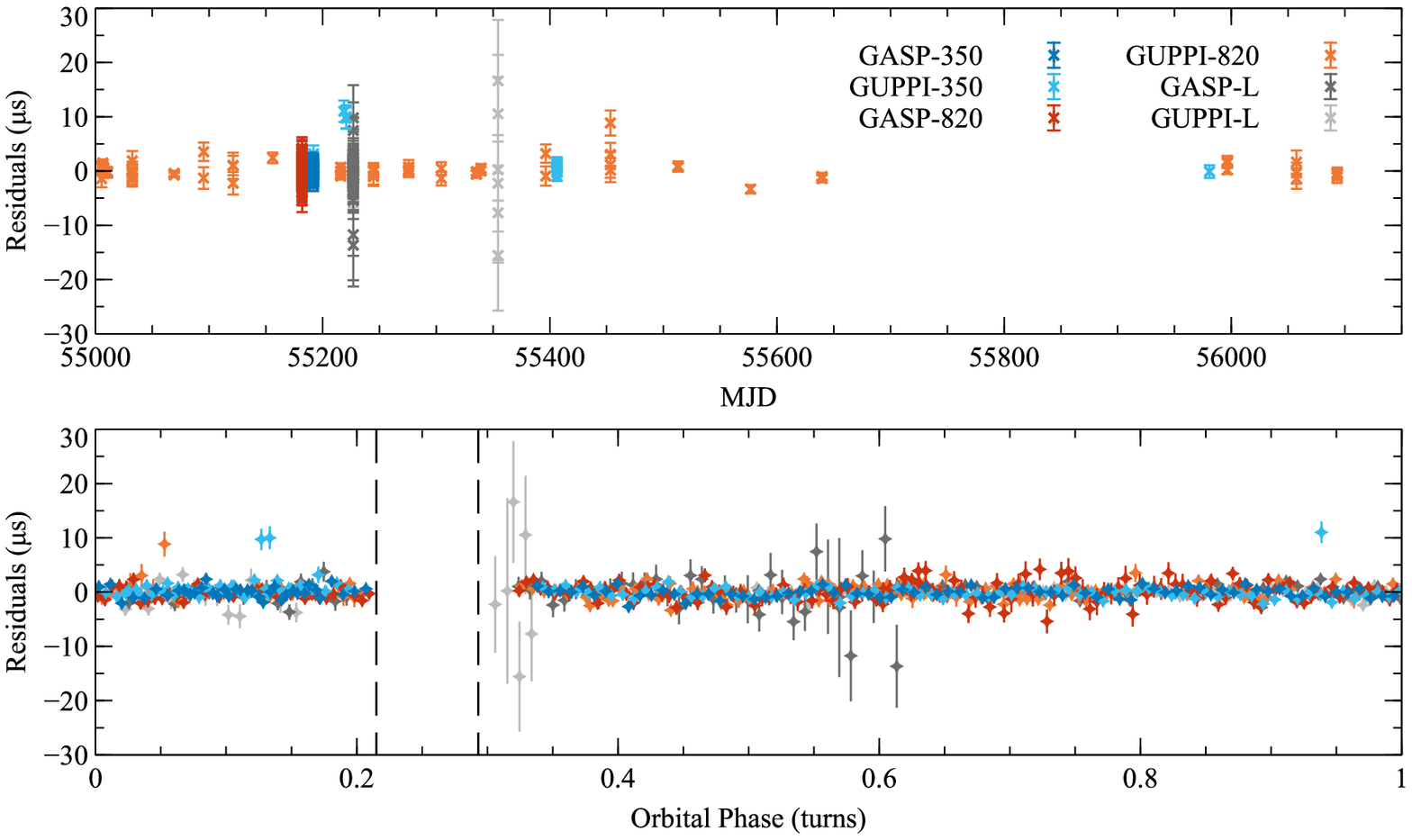}
    \caption{Timing residuals as a function of time and orbital phase for the radio timing solution in \autoref{tbl:par_file_pars}. Dashed lines show the approximate location of the eclipse, between phases 0.215-0.293. The eclipse and its surroundings have been excluded.}
    \label{fig:res_v_mjd_and_ophase}
\end{figure*}

A timing analysis was performed using the TEMPO\footnote{http://tempo.sourceforge.net} software package.
From an initial timing parameter file, we found a model for the TOAs over our data span which gave a phase-connected solution wherein every rotation of the neutron star is accounted for. The predicted pulse TOAs are subtracted from the measured TOAs to produce residuals. \autoref{tbl:par_file_pars} gives the final timing solution found for J2256--1024 using the JPL DE436 solar system ephemeris. The residuals from this model are shown in \autoref{fig:res_v_mjd_and_ophase}. Values derived from these parameters, such as the characteristic age of the pulsar, are then given in \autoref{tbl:derived_parameters}.

\begin{table*}
    \begin{threeparttable}
    \begin{tabular}{lll}
    	\hline
    	\hline
        \multicolumn{3}{c}{Radio Timing Solution} \\
        \hline
        \hline
        \multicolumn{3}{c}{Pulsar Parameters} \\
        \hline
        TEMPO Shorthand & Description & {Value} \\
        \hline
        RAJ & J2000 Right ascension (hh:mm:ss) & {\ra{22;56;56.39294 \pm 0.00007}} \\
        DECJ & J2000 Declination (dd:mm:ss) & {\dec{-10;24;34.385 \pm 0.003}} \\
        F0 & Rotational frequency (\si{s^{-1}}) & \num{435.8187550969386 \pm 0.000000000004} \\
        F1 & 1st time derivative of the rotational frequency (\si{s^{-2}}) & \num{-2.15646 \pm 0.00018E-15} \\
        DM & Dispersion measure (\si{cm^{-3}\parsec}) & \num{13.776020 \pm 0.000003} \\
        PX\tnote{*} & Parallax (\si{\mas}) & \num{0.48 \pm 0.15} \\
        PMRA\tnote{*} & Proper motion in right ascension (\si{\masyr}) & \num[separate-uncertainty=true]{3.2\pm 1.1} \\
        PMDEC\tnote{*} & Proper motion in declination (\si{\masyr}) & \num[separate-uncertainty=true]{-8.5 \pm 2.7} \\
        \hline
        \multicolumn{3}{c}{Binary Parameters} \\
        \hline
        A1 & Projected semi-major axis of pulsar orbit (\si{\lightsecond}) & \num{0.08296575 \pm 0.00000005} \\
        E & Eccentricity of orbit & 0 \\
        T0 & Epoch of periastron (MJD) & \num{55548.92435263 \pm 0.00000011} \\
        OM & Longitude of periastron (\si{\degree}) & 0 \\
        FB0 & Orbital frequency (\si{s^{-1}}) & \num{5.4368334547 \pm 0.0000000016E-5} \\
        \hline

        \multicolumn{3}{c}{Assumptions and Model Parameters} \\
		\hline
		PEPOCH & Epoch of period determination (MJD) & {55549} \\
		EPHEM & Solar system ephemeris used & {DE436} \\
		CLK & Clock correction used & {TT(BIPM)} \\
		SOLARN0 & Proportionality constant for solar wind model (\si{electrons.cm^{-3}} at \SI{1}{\au}) & {5.00} \\
		BINARY & Binary model used & {BTX} \\
		\hline
		\multicolumn{3}{c}{Data Statistics} \\
		\hline
		START & Start of data span (MJD) & 55005.385 \\
		FINISH & End of data span (MJD) & 56093.266 \\
		NTOA & Number of TOAs & 773 \\
		TRES & RMS timing residual (\si{\micro\second}) & 0.99 \\
		\hline
		\hline
        \multicolumn{3}{c}{Gamma-ray Timing} \\
		\hline
		\hline
		\rule{0pt}{10pt}PMRA & Proper motion in right ascension (\si{\masyr})& \begin{math}4.0^{+2.9}_{-2.8}\end{math} \\[3pt]
        PMDEC & Proper motion in declination (\si{\masyr}) & \begin{math}-7.6^{+6.8}_{-7.1}\end{math} \\[3pt]
        FB1 & 1st time derivative of the orbital frequency (\si{\per\second\squared}) & \begin{math}-4.2^{+0.7}_{-0.7} \times 10^{-21}\end{math} \\

	\end{tabular}
	\begin{tablenotes}
		\item{*} tentative - see discussion in \autoref{subsec:px_pm}
	\end{tablenotes}
	\end{threeparttable}
	\caption{\label{tbl:par_file_pars} Radio timing solution for J2256--1024, plus additional parameters resulting from the gamma-ray analysis described in \autoref{subsec:gamma_stuff}. Quantities in parentheses indicate the uncertainty in the last digits, e.g. \num{5.55 \pm 0.11} would correspond to \num[separate-uncertainty=true]{5.55 \pm 0.11}. In the radio section, errors shown are as output by TEMPO and F1 has not been corrected for galactic acceleration. The gamma ray section shows median values with their \SI{68}{\percent} confidence limits.}
\end{table*}



\autoref{fig:res_v_mjd_and_ophase} shows some remaining scatter in the residuals. A large scatter in residuals with large error bars within a single epoch (such as GUPPI-L on MJD 55226) are due to low S/N and the pulse only appearing in a small subsection of the band. There are also some epochs with a small scatter and small error bars which appear to be outliers. Such epochs were investigated but no reason was found to exclude them. We were unable to fit for DM variations across the data span as multi-frequency epochs are sparse; both in-band DM determination and dividing the GUPPI bandwidth into a number of sub-bands were also unsuccessful. This is a black widow system, where there is likely a varying DM due to changing amounts of extra material in the system, thus some residual scatter is expected.
Given these factors the reduced chi squared found, \begin{math}\chi^2_{\rm red}=1.41\end{math}, seems reasonable. There was no reason to suspect uncertainties from a particular instrumental configuration were undervalued and so no error-raising factors, such as EFAC or EQUAD, were used. As \begin{math}\chi^2_{\rm red}>1\end{math} for the final timing solution, the uncertainties quoted in \autoref{tbl:par_file_pars} and \autoref{tbl:derived_parameters} will be slightly undervalued.

Over the \SI{3}{\yearunit} span small variations in the dispersion measure are expected, due to the changing path between Earth and J2256--1024 and shifting material in the interstellar medium. Unfortunately, as shown in \autoref{fig:ophase_v_mjd}, epochs with multi-frequency data are rare. Dispersion being a frequency-dependent effect, this meant any DM variations present could not be measured. 

\subsection{Parallax and Proper Motion}
\label{subsec:px_pm}
We are presenting the parallax as a tentative measurement. Its inclusion had no visible effect on the residuals, gave only a small statistical improvement to the fit - improving the weighted root mean square (RMS) residual by \SI{0.006}{\micro\second} (\SI{0.6}{\percent}) - and gave a less than 3-sigma significance [\begin{math}\nicefrac{\text{value}}{(\text{uncertainty}~ \times \sqrt{\chi^2_{\rm red}})}=2.83\end{math}]. However, parallax manifests itself in residuals as a repeating signal rather than a growing one and folding our residuals into a period of one year shows decent coverage. Furthermore, an F-test comparing models without and with parallax gave a p-value of 0.0031 that the improvement in the $\chi^2$, upon the inclusion of parallax, was due to chance. In addition, the distance to J2256--1024 derived from the parallax measurement is compatible with that found from combining the measured DM with the YMW16 model for the Galactic electron density distribution~\citep{Yao2016}, both of which are given in \autoref{tbl:derived_parameters}. We have not attempted to correct the parallax for Lutz-Kelker bias (e.g. \cite{Verbiest2010a}).

We are also presenting proper motion in right ascension and declination as a tentative measurement and using these values to provide a \SI{95}{\percent} upper limit on the proper motion (\autoref{tbl:derived_parameters}). The radio data span is relatively short and has some notable gaps in coverage, making determining the proper motion challenging. Also, statistical improvements to the fit are less strong: PMRA and PMDEC have significances of \num{2.6} and \num{2.7} respectively when incorporating the \begin{math}\sqrt{\chi^2_{\rm red}}\end{math} of the fit as above, including proper motion improves the weighted RMS residual by \SI{0.005}{\micro\second} (\SI{0.5}{\percent}), and an F-test gives a p-value of \num{0.057}. However, radio-data fits for the proper motion gives values consistent with the gamma-ray timing analysis of {\em Fermi} LAT data described in \autoref{subsec:gamma_stuff}. Some parameters from the radio timing solution were held fixed in the gamma-ray analysis, but otherwise these two data sets and analyses are independent. Thus, proper motion was included in the radio timing solution.

When correcting \(\dot{f}\) for the effects of Galactic acceleration, the Shklovskii component~\citep{Shklovskii1970} was not included. However, the proper motion upper limit provides a limit on the size of the Shklovskii correction to \(\dot{f}\), which is also listed in \autoref{tbl:derived_parameters}.

\subsection{{\em Fermi} Timing Analysis}
\label{subsec:gamma_stuff}

In order to attempt to measure long-term timing parameters such as proper motion and a possible orbital period derivative, we conducted a single-photon timing analysis using all available {\em Fermi} LAT `Source Class' events on J2256--1024.  We downloaded archived Pass 8 R3 events \citep{P8} above 100\,MeV and within 15\degr\ of the pulsar between 2008 August 5 and 2019 April 15.  We used an unbinned Markov Chain Monte Carlo (MCMC) analysis with the {\tt event\_optimize} code in {\tt PINT} \citep{PINT} and the {\tt emcee} sampler \citep{emcee}.  The analysis is based on the Maximum Likelihood technique described in \cite{2PC} and \cite{pc15}.

We used a noise-less pulse template based on five gaussians, determined via iterating the MCMC timing analysis and improving the template made of the summed photons each time.  We weighted each event using the default scheme for LAT photons in {\tt event\_optimize}, which is based on the spectrum of a typical gamma-ray pulsar, the known position of J2256--1024, and the energy-dependent point source response function of the {\em Fermi} LAT.

\begin{figure}
	\includegraphics[]{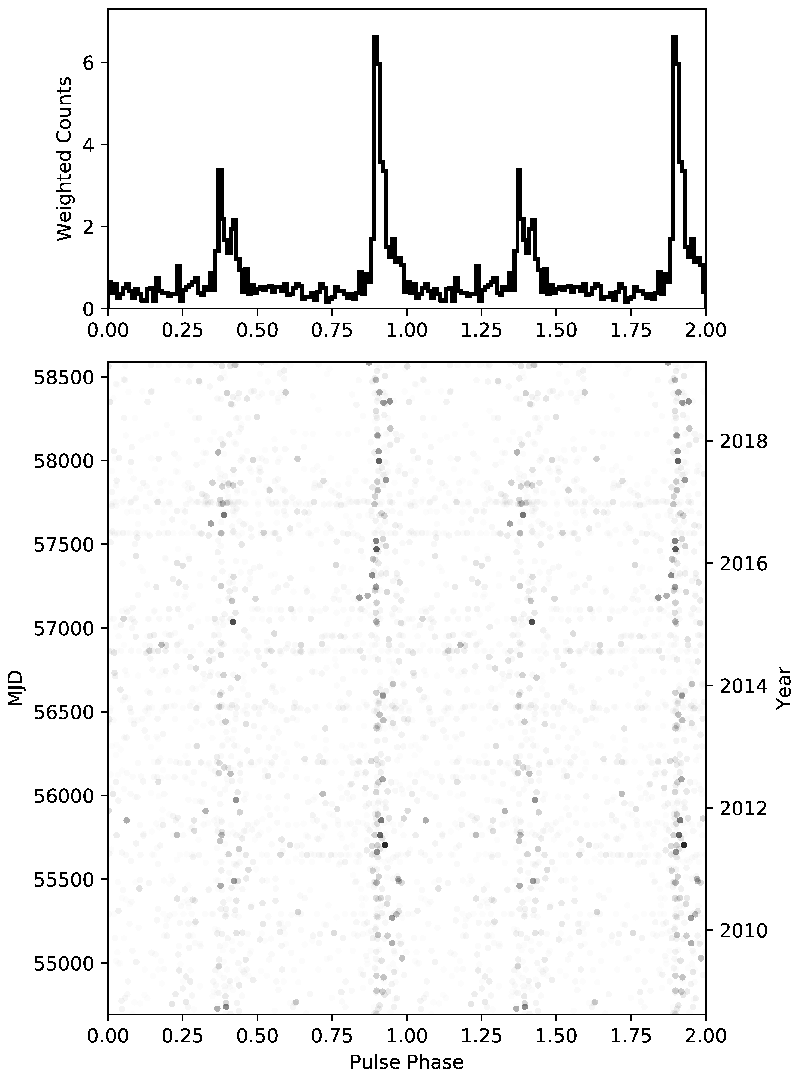}
    \caption{The summed gamma-ray pulse profile (top) and time-vs-phase diagram (bottom) of the weighted {\em Fermi} LAT events towards J2256--1024.}
    \label{fig:Fermi_photon}
\end{figure}

For this timing analysis, we fixed the radio ephemeris and only fit for the pulsar's spin frequency, first spin frequency derivative (for both of which we measured a value consistent with the radio ephemeris), parallax, the proper motion of the pulsar in both RA and DEC, and a first orbital frequency derivative.  The median values and 68\% confidence limits for these parameters are listed in \autoref{tbl:par_file_pars}. No significant constraint on the parallax was found. As noted above, values found for the proper motion are consistent with the radio analysis. A ``phasogram'' and best gamma-ray pulse profile from this analysis are shown in Figure~\ref{fig:Fermi_photon}.

\subsection{Derived Quantities}
\label{subsec:derived_quantities}

\begin{table*}
	\begin{tabular}{llll}
		\hline
		\multicolumn{4}{c}{(i) Using TEMPO Fit Parameters} \\
		\hline
		Symbol & Description & \multicolumn{2}{l}{Value} \\
		\hline
        \begin{math}l\end{math} & Galactic Longitude & \multicolumn{2}{l}{\ang{59.23}} \\
        \begin{math}b\end{math} & Galactic Latitude & \multicolumn{2}{l}{\ang{-58.29}} \\
		\begin{math}f_{\rm m}\left( m_{\rm p},m_{\rm c}\right)\end{math} & Mass function & \multicolumn{2}{l}{\SI{0.00001353003 \pm 0.00000000003}{\msun}}\\
		\begin{math}d_{\rm DM}\end{math} & Distance, inferred from DM (YMW16) & \multicolumn{2}{l}{\begin{math}1.3^{+0.4}_{-0.3}\end{math}~\si{\kilo\parsec}} \\ 
		\begin{math}d_{\rm DM}\end{math} & Distance, inferred from DM (NE2001) & \multicolumn{2}{l}{\begin{math}0.65^{+0.11}_{-0.10}\end{math}~\si{\kilo\parsec}} \\ 
		\begin{math}d_{\rm px}\end{math} & Distance, from PX & \multicolumn{2}{l}{\SI{2.0 \pm 0.6}{\kilo\parsec}} \\
		\begin{math}\dot f\end{math} & F1, corrected for Galactic acceleration & \multicolumn{2}{l}{\SI{-2.073 \pm 0.009 E-15}{s^{-2}}} \\
		\begin{math}\mu\end{math} & \SI{95}{\percent} Upper Limit on Proper Motion & \multicolumn{2}{l}{\SI{14}{\masyr}} \\
		\begin{math}\Delta\dot f _{\rm Shk}\end{math} & Shklovskii Correction Upper Limit & \multicolumn{2}{l}{\SI{-2.8 E-16}{s^{-2}}} \\
		\hline
		\multicolumn{4}{c}{(ii) Assuming a Pulsar with Moment of Inertia \SI{E+45}{\g\cm\squared}} \\
		\hline 
		\begin{math}\tau\end{math} & Characteristic Age & \multicolumn{2}{l}{\SI{3.3}{\giga\yearunit}} \\
		\begin{math}E_{\rm rot}\end{math} & Rotational Kinetic Energy & \multicolumn{2}{l}{\SI{3.7 E51}{\erg}} \\
		\begin{math}\dot {E}_{\rm rot}\end{math} & Rate of Change of Rotational Kinetic Energy & \multicolumn{2}{l}{\SI{-3.6 E34}{\erg ~\s^{-1}}} \\
		\begin{math}B_{\rm min}\end{math} & Minimum Surface Magnetic Field & \multicolumn{2}{l}{\SI{1.6 E8}{\gauss}} \\
			
		\hline
		\multicolumn{4}{c}{(iii) Assuming a Pulsar Mass of \SI{1.4}{\msun} and \begin{math}i=\ang{90}\end{math}} \\
		\hline
		\begin{math}m_{\rm c}^{\rm min}\end{math} & Minimum Companion Mass & \multicolumn{2}{l}{\SI{0.030248740 \pm 0.000000019}{\msun}} \\
		\begin{math}a\end{math} & Pulsar-Companion Separation &  \SI{3.922863 \pm 0.000003}{\lightsecond} & \SI{1.6904446 \pm 0.0000015}{\rsun}\\
		\begin{math}R_{\rm L}^{\rm c}\end{math} & Companion's Effective Roche Lobe Radius & \SI{0.510 \pm 0.005}{\lightsecond} & \SI{0.220 \pm 0.002}{\rsun} \\ 
			
		\hline
		\multicolumn{4}{c}{(iv) Incorporating Inclination Angle of \ang{68 \pm 11} from \citet{1302.1790}} \\ 
		\hline
		\begin{math}m_{\rm c}\end{math} & Companion Mass & \multicolumn{2}{l}{\SI{0.0327 \pm 0.0010}{\msun}} \\
		\begin{math}a\end{math} & Pulsar-Companion Separation & \SI{3.9 \pm 0.3}{\lightsecond} & \SI{1.69 \pm 0.14}{\rsun}\\ 
		\begin{math}R_{\rm L}^{\rm c}\end{math} & Companion's Effective Roche Lobe Radius & \SI{0.52 \pm 0.05}{\lightsecond} & \SI{0.23 \pm 0.02}{\rsun} \\ 
		\hline
	\end{tabular}
		
	\caption{\label{tbl:derived_parameters} Derived parameters under various assumptions. Uncertainties shown have been propagated from those output by TEMPO. In (i) for the YMW16 DM distance, the uncertainty stems from assuming a \SI{20}{\percent} error in the DM measurement as was standard in the NE2001 model~\citep{astro-ph/0207156}. The Galactic acceleration correction in $\dot f$ does not include the the Shklovskii component; an upper limit of that correction is also listed. In (iii) and (iv) the Roche lobe radius was calculated using the \citet{1983ApJ...268..368E} approximation. The given Roche lobe uncertainty in (iii) is the quoted \SI{1}{\percent} maximum disagreement between the approximation and numerical integration; in (iv) the uncertainty was propagated from its dominant source - the pulsar-companion separation.}
		
\end{table*}


\autoref{tbl:derived_parameters} lists further parameters describing the pulsar, the system, and its companion. These were derived using the timing parameters in the final radio timing solution given by TEMPO in \autoref{tbl:par_file_pars}, making a series of assumptions, incorporating results from independent models, and using values from the companion detection paper.

\subsection{Galactic Acceleration correction}
The observed \begin{math}\dot{f}\end{math} in \autoref{tbl:par_file_pars} differs from the intrinsic value due to a Doppler shift caused by the relative accelerations of the pulsar system and the Solar System Barycentre within the Milky Way. The transverse component (the Shklovskii effect) of the correction was not applied to \begin{math}\dot{f}\end{math} but an upper limit is given in \autoref{tbl:derived_parameters}. The reported \begin{math}\dot{f}\end{math}, and any values derived from it, should be considered with this in mind.

The effect due to the line-of-sight component has been corrected for, following \citet{Nice1995}. To find the acceleration towards the plane we use the \citet{Kuijken1989} model for the mass distribution in the Galactic disk (with a local mass density of \begin{math}\rho = \SI{1E-2}{\msun\per\cubic\parsec}\end{math} and a total disk column density of \begin{math}\Sigma = \SI{46}{\msun\per\square\parsec}\end{math}). To find the acceleration due to the differing Galactic rotations we assume a flat rotation curve and use the \citet{Reid2014} values for the distance to the Galactic center, \begin{math}R_0 = \SI{8.34 \pm 0.16}{\kilo\parsec}\end{math}, and its rotational velocity, \begin{math}\Theta_0 = \SI{240 \pm 8}{\kilo\meter\per\second}\end{math}.

For these corrections we used the YMW16 distance derived from the DM as our parallax measurement is tentative; as noted, these inferred distances are similar in any case, so resulting variations in the final values are small or negligible. The corrected \begin{math}\dot{f}\end{math} and derived quantities are reported in \autoref{tbl:derived_parameters}.

\subsection{Companion Detection Paper}
\label{subsec:optical_companion_paper}
\citet{1302.1790} reported a detection of J2256--1024's companion star in the optical. Using a preliminary timing solution and light curve fitting, they found an intermediate inclination angle for the system of \ang{68 \pm 11} and that the companion was under-filling its Roche Lobe with a filling factor\footnote{the ratio of the companion's radius to its effective Roche lobe radius} of \num{0.4 \pm 0.2}. However, \citet{1302.1790} used a distance of \SI{0.65}{\kilo\parsec}, derived from the DM using the NE2001 model. This is much smaller than the distance derived through either the parallax measurement or through using the DM with the YMW16 model. \citet{1302.1790} note that holding the filling factor fixed at 1 requires a distance of \SI{1.5}{\kilo\parsec} to produce their measured fluxes. This is consistent with our reported YMW16 and parallax distances; thus we infer that the companion is filling (or even over-filling) its Roche lobe. \citet{1302.1790} do not mention whether using this larger distance affected their inclination angle.

\begin{figure*}
    \includegraphics[]{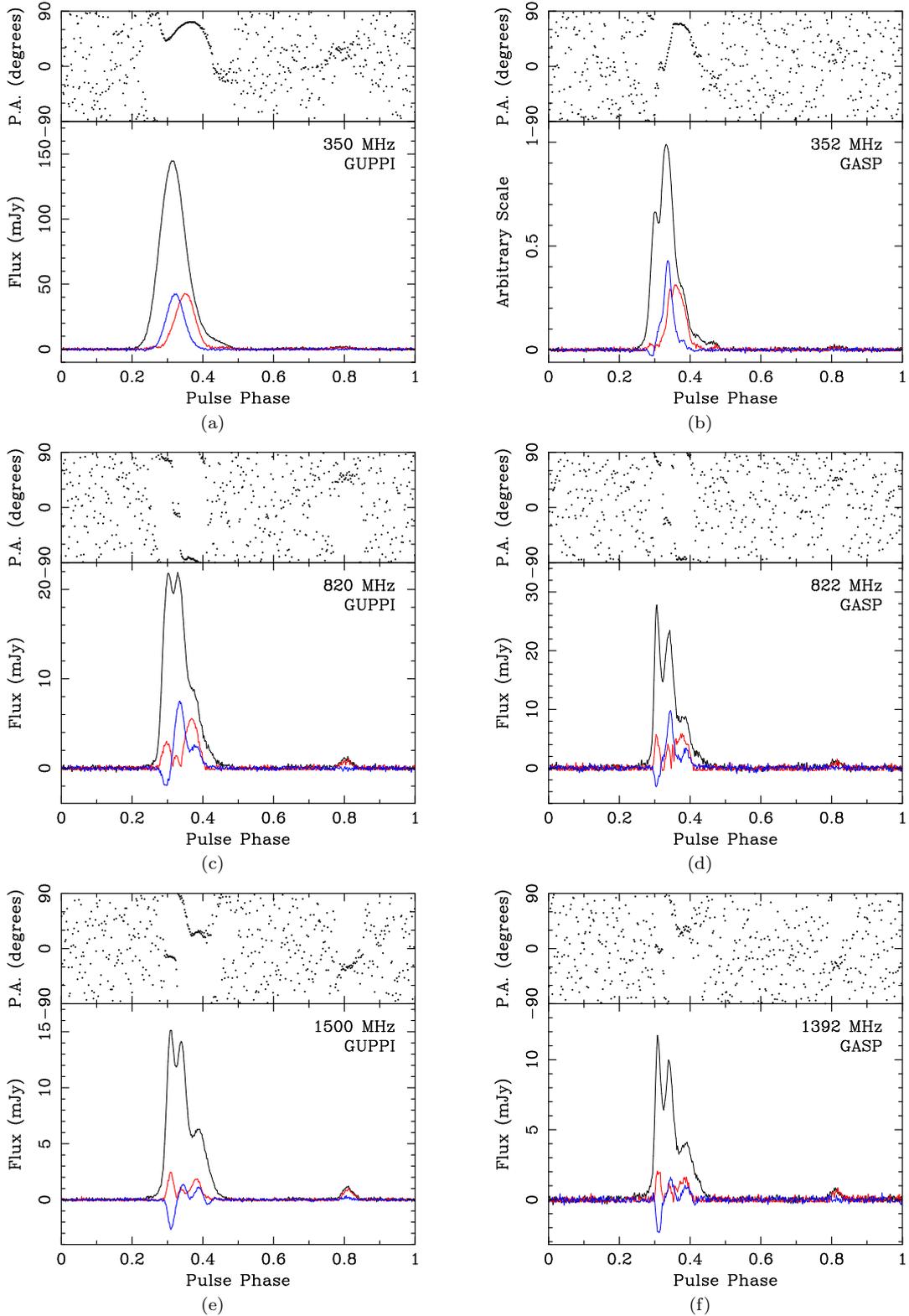}
    \caption{\label{fig:pol_profiles}Polarization profiles for the backend-frequency combinations shown. In each figure the upper plot shows the position angle, the lower plot shows the total intensity (black), linear polarization (red), and circular polarization (blue). There are 512 pulse phase bins in each profile. (a) MJD 55191. Incoherently dedispersed and summed over 100~MHz and 5.02~hr. (b) MJD 55191. Coherently dedispersed and summed over 24~MHz and 5.67~hr. Note this observation could only be partially calibrated, thus the profile is shown on an arbitrary scale.(c) MJD 55181. Incoherently dedispersed and summed over 200~MHz and 4.64~hr. (d) MJD 55181. Coherently dedispersed and summed over 64~MHz and 5.34~hr. (e) MJD 55226. Incoherently dedispersed and summed over 800~MHz and 3.41~hr. (f) MJD 55226.  Coherently dedispersed and summed over 84~MHz and 5.74~hr.}
\end{figure*}


\section{Polarization Profiles}
\label{sec:pol_profiles}

\autoref{fig:pol_profiles} presents polarization profiles for J2256--1024 at \num{350}, \num{850} and \SI{1500}{MHz} made from observations taken concurrently with GASP and GUPPI on MJDs 55191, 55181, and 55226 respectively. The profiles have been rotated by 0.3 in pulse phase for easier viewing. Profiles at each epoch have been RM-corrected with the RM measured from the GUPPI observation at that epoch. These RMs - \SI{13.4 \pm 0.5}{rad/m^2} on MJD 55181 as \SI{820}{\MHz}, \SI{15.04 \pm 0.05}{rad/m^2} on MJD 55191 as \SI{350}{\MHz}, and \SI{14.2 \pm 0.8}{rad/m^2} on MJD 55226 at L-band respectively - include both intrinsic and ionospheric contributions. Due to the absence of a fluxcal, the GASP \SI{350}{\MHz} observation on MJD 55191 underwent the less robust calibration procedure described in \autoref{sec:obs_data} and its plot therefore shows relative flux. Other differences between GASP and GUPPI profiles are likely due to: (a) the different dedispersion processes; (b) GASP bands being a small subset of the GUPPI bands.

In the two higher frequency bands, an interpulse is clearly visible. It is less visible in the \SI{350}{\MHz} plots but this is partially due to scaling; interestingly the flux density of the interpulse appears to remain approximately constant in all profiles. \citet{Dai2015} and \citet{Bhat2018} found very complex profile and polarization frequency evolution in many MSPs, including variations in the spectral index across the pulse profile. The sparsity of calibrated, multi-frequency data prevents us drawing similar conclusions about J2256--1024. By comparing the profiles at different frequencies we also see some profile evolution; in the main double peak, the intensity of the earlier peak increases with respect to the latter as frequency increases.

\section{Scintillation and Spectrum}
\label{sec:dyn_spec}

\begin{table}
\centering
	\begin{tabularx}{0.9\linewidth}{lXXXX}
		\hline
		\multirow{2}{*}{Backend} & Central Frequency & Mean Flux Density & $\Delta t_{\rm DIFF}$ & $\Delta \nu_{\rm DIFF}$ \\ 
		& \si{MHz} & \si{\milli\jansky} & \si{\second} & \si{\MHz} \\ 
		\hline 
		GUPPI & 350 & \num{13} & \num{1180 \pm 40} & \num{0.41 \pm 0.02} \\
		\hline
		GASP & 822 & \num{1.7} & & \\ 
		GUPPI & 820  & \num{1.9 \pm 0.9} & \num{3100 \pm 500} & \num{8.9 \pm 1.8} \\
		\hline
		GASP & 1392 & \num{0.73} & & \\
		GUPPI & 1500 & \num{1.2} & - & - \\
		\hline
	\end{tabularx}
	
	\caption{\label{tbl:flux_freq} Mean flux densities, plus decorrelation time scales and bandwidths due to diffractive interstellar scintillation. Mean flux densities were calculated from the observations which form the profiles in \autoref{fig:pol_profiles}, excepting the GUPPI 820 MHz value. This is the mean value from all calibrated GUPPI observations at 820 MHz, and the uncertainty shown is the standard deviation of these measurements.
	Scintillation values given were measured from the GUPPI observations on MJDs 55181 and 55191. No scintles could be resolved at L-band. No other observations were of long enough duration for scintles to be resolved and measured.}
\end{table}

\begin{figure}
    \includegraphics[]{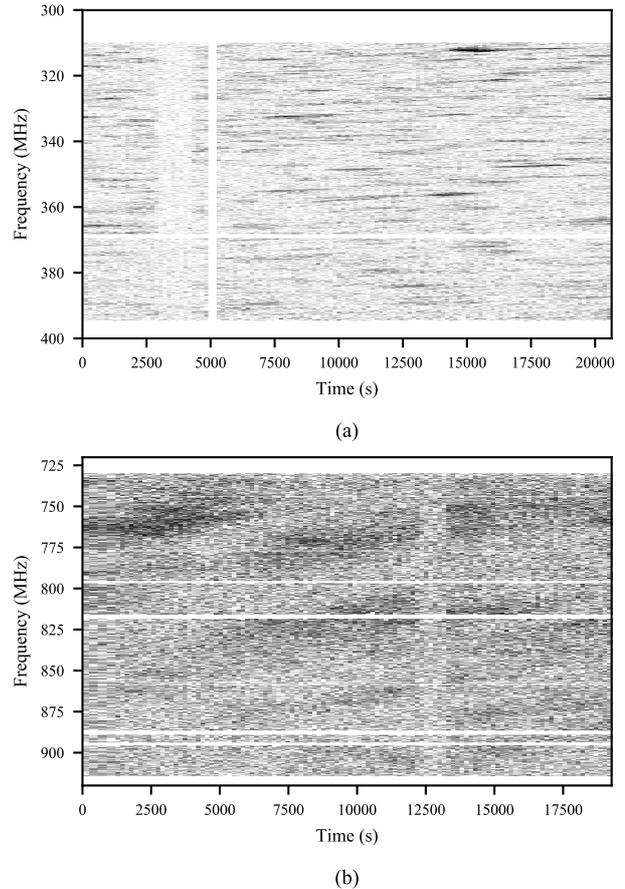}
    \caption{Dynamic spectra formed from flux calibrated GUPPI observations (a) at 350 MHz on MJD 55191, and (b) at 820 MHz on MJD 55181. White lines occur where frequency channels and sub-integrations were zero-weighted due to RFI.}
    \label{fig:dynspec}
\end{figure}


From the profiles given in \autoref{fig:pol_profiles} we also compute a mean flux density for each fully calibrated backend-frequency combination (with the exception of GUPPI-820 MHz as discussed below); these are given in \autoref{tbl:flux_freq}. Pulsar fluxes are known to vary in time due to diffractive and refractive interstellar scintillation (DISS and RISS)~\citep{Narayan1992}. DISS decorrelation bandwidths and time scales were measured using \texttt{PyPulse}~\citep{Lam2017} by forming a 2D autocorrelation of the dynamic spectra, then fitting a rotated 2D Gaussian. Uncertainties in $\Delta t_d$ and $\Delta \nu_d$ were computed assuming the dominant source of error is the finite number of scintillation features within the dynamic spectra, as per \citet{Cordes1986}. These values are given in \autoref{tbl:flux_freq} and \autoref{fig:dynspec} shows dynamic spectra for the two epochs with scintillation features. Scintles could not be resolved at L-band; likely the decorrelation time scale at that frequency is longer than the duration of the observation. Likewise, no other calibrated observations were long enough for scintillation features to be resolved. 
Using $\Delta t_{\rm DISS}$ to estimate the RISS timescale via $\Delta t_{\rm RISS} = \Delta t_{\rm DISS} \times \nicefrac{\nu}{\Delta \nu_{\rm DISS}}$, where $\nu$ is the observing frequency~\citep{2004hpa..book.....L}, the refractive timescale at 820 MHz is approximately 3.3 days. This is much smaller than the time between epochs for most of our calibrated 820 MHz data. Therefore, we computed mean flux densities for each calibrated 820 MHz observation; the value given in \autoref{tbl:flux_freq} is the mean and standard deviation of these measurements. Unfortunately this process could not be repeated at 350 MHz as our data only contain one calibrated GUPPI observation.

Assuming the GUPPI 820 MHz percentage error applies to the mean flux density at all frequencies, and performing a simple linear fit on a logarithmic plot, we calculate a spectral index of \num{-1.8 \pm 0.5}. We caution that this measurement is not robust as measurements at 350 MHz and L-band are each based on a single epoch; this is particularly harmful at L-band where scintillation timescales and bandwidth will be larger.

\citet{Jankowski2018} studied the pulsar population as a whole with a sample of 441 pulsars, and found, of those pulsars whose spectra followed a simple power law, a weighted mean spectral index of \num{-1.60 \pm 0.03}. There has been some suggestion that the population of gamma-ray millisecond pulsars tend to have steeper spectra in \citet{Kuniyoshi2015} and \citet{Frail2016}. Both papers caution that this may be due to biases; however, later works (such as \citet{Bassa2017} and \citet{Kaur2019}) add evidence for this theory. If true then J2256--1024 has a comparatively shallow spectrum within that subset.

\section{Eclipse Analysis}
\label{sec:eclipse}


\begin{figure}
	\includegraphics[]{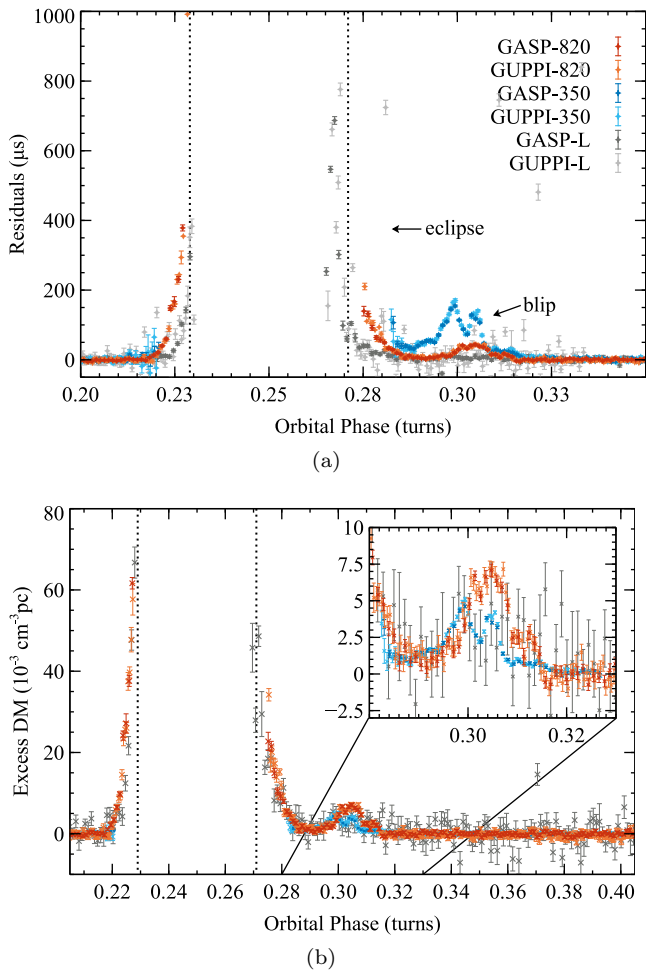}
    \caption{Higher time resolution data for all observed eclipses. The \SI{820}{\MHz}, \SI{350}{\MHz}, and L-band data were taken on MJDs 55181, 55191, and 55226 respectively. \label{fig:higher_res} (a) Timing residuals for the radio timing solution in \autoref{tbl:par_file_pars}. Dotted vertical lines (also shown in (b)) indicate the extent of the companion's Roche lobe assuming a \SI{1.4}{\msun} pulsar, a \ang{90} inclination angle, and that it is positioned symmetrically about conjunction. However, it should be noted that \citet{1302.1790} found an inclination angle of \ang{68 \pm 11} based on light curve modelling, in which case the companion's Roche lobe does not intersect the line-of-sight. (b) Excess dispersion measure in the eclipse region. The inset shows the ``blip" discussed in \autoref{subsec:blip_dm} in more detail. GUPPI L-band data has been excluded for clarity.}

\end{figure}


Only three epochs cover the eclipse, with one at each frequency. TEMPO was used to generate residuals by using the higher-time-resolution TOAs described in \autoref{subsec:data_reduction_eclipse} and holding the radio timing solution of \autoref{tbl:par_file_pars} fixed. These higher-time-resolution residuals are shown in \autoref{fig:higher_res}a with the companion's inferior conjunction marked at 0.25 in orbital phase. 

The eclipse shows some asymmetry, with an ingress a little sharper than its egress. Eclipse asymmetry is typical of black widow systems and was noted in the original B1957+20 discovery paper~\citep{1988Natur.333..237F}. After the eclipse there is a group of delayed pulses - for lack of a better term, a ``blip" - which appears in both the \SI{820}{\MHz} observation on MJD 55181 and 10 days later at \SI{350}{\\MHz}; this is discussed later.
Residuals in \autoref{fig:higher_res}a were scaled by a factor of \begin{math}\nicefrac{\nu^2}{K}\end{math}, where \begin{math}K=\SI{4.148808E3}{\MHz\squared\centi\metre\cubed\second\per\parsec}\end{math}, to form an ``Excess DM" which is then plotted in \autoref{fig:higher_res}b. 

The duration and shape of the eclipse at different frequencies, shown in \autoref{fig:higher_res}b, confirms that the eclipse follows the normal dispersive \begin{math}\nicefrac{1}{\nu^2}\end{math} frequency dependence. There is some hint the \SI{350}{\MHz} eclipse exit may be sharper than that at \SI{820}{\MHz}, but as these observations were taken 10 days apart this may be due to real changes in the amount and/or distribution of eclipsing material present. By inspecting the plot the eclipse is approximately from phase \num{0.215} to \num{0.293} but determining the ``end'' of the eclipse is somewhat difficult to determine as the excess DM does not return to baseline between the eclipse and the blip. In \autoref{fig:higher_res}b we see the asymmetry of the eclipse more clearly and marked on the plot is the projected size of the companion's Roche lobe, if the orbit was perfectly edge on, centred at \num{0.25} in orbital phase. The pulse delays and excess material in the path clearly both start and end past the extent of the companion's Roche lobe. This agrees with the classic picture of a black widow where the material ``blown off'' the companion forms a cloud of some kind around and near it (likely with some kind of cometary tail~\citep{Rasio1989, AlessandroRidolfi2012}), and this cloud of material causes the eclipses in addition to the companion itself. For the more intermediate inclination angle of \ang{68 \pm 11} found by \citet{1302.1790} the companion's Roche lobe would not intersect the line-of-sight at all and the cloud would be the sole cause of the eclipse.

\subsection{The post-eclipse blip feature}
\label{subsec:blip_dm}

From \autoref{fig:higher_res} the blip is not visible in the L-band observation. It appears to be a distinct feature separate from the eclipse tail, yet, inspecting the inset, the excess DM does not return to baseline in between egress and the blip. The blip appears in data taken with both backends both at \SI{350}{\MHz} on MJD 55191 and at \SI{820}{\MHz} on MJD 55181. The only two other observations which sample this region of orbital phase were performed at L-band and some time later - MJDs 55226 and 55345; blips were not seen in either of the observations. There is no data corruption or discernible errors in the 55181 and 55191 observations. 
Therefore, we are confident that the blip is a real feature and likely due to some clump of material. Given that the excess DM does not fall back to zero before the blip's occurrence, it may well be a clump within the comet-like tail or cloud coming off the companion. 

The inset in \autoref{fig:higher_res}b shows a close-up of the blip region. The blips detected at \num{350} and \SI{820}{\MHz} are not consistent with each other. This suggests several possibilities: the separate blips could be due to separate clumps; the blips could be due to the same clump of material, which then changed its morphology over the intervening 10 days between observations; and/or the differences are due to probing the clump at different frequencies. Without more blip incidents we can only speculate.

It is also clear from \autoref{fig:higher_res}b that if a similar clump were present on MJD 55226 we would not have been able to detect it at L-band. Clumps such as these may be rare and the observations on MJDs 55181 and 55191 fortuitously timed but, given we found evidence of clumps on the only two occasions when this region of orbital phase was sampled with a frequency likely to detect them, it is likely clumps are a common occurrence.

This is supported by off-eclipse dispersive delay events seen in other black widow and redback systems: A blip is visible in Fig.~1 of \citet{Main2018}, a recent paper on B1957+20; \citet{Deneva2016} see ``mini-eclipses'' in J1048+2339; \citet{Archibald2009a} note a blip in J1023+0038 due to large variations in DM when exiting the eclipse; variable dispersion measures are frequently reported for PSR B1744--24A (also known as Ter5~A), e.g. \citet{Bilous2018}; and \citet{Polzin2018} see ``significant deviations from the out-of-eclipse electron column density" in J1810+1744. Our blips seem to be part of the eclipsing cometary tail; given \citet{Stappers1996} found indications of variable structure in J2051--0827's eclipsing material, it seems reasonable structure would also be present further out in the tail.

\begin{landscape}
    \begin{figure}
    	\includegraphics[]{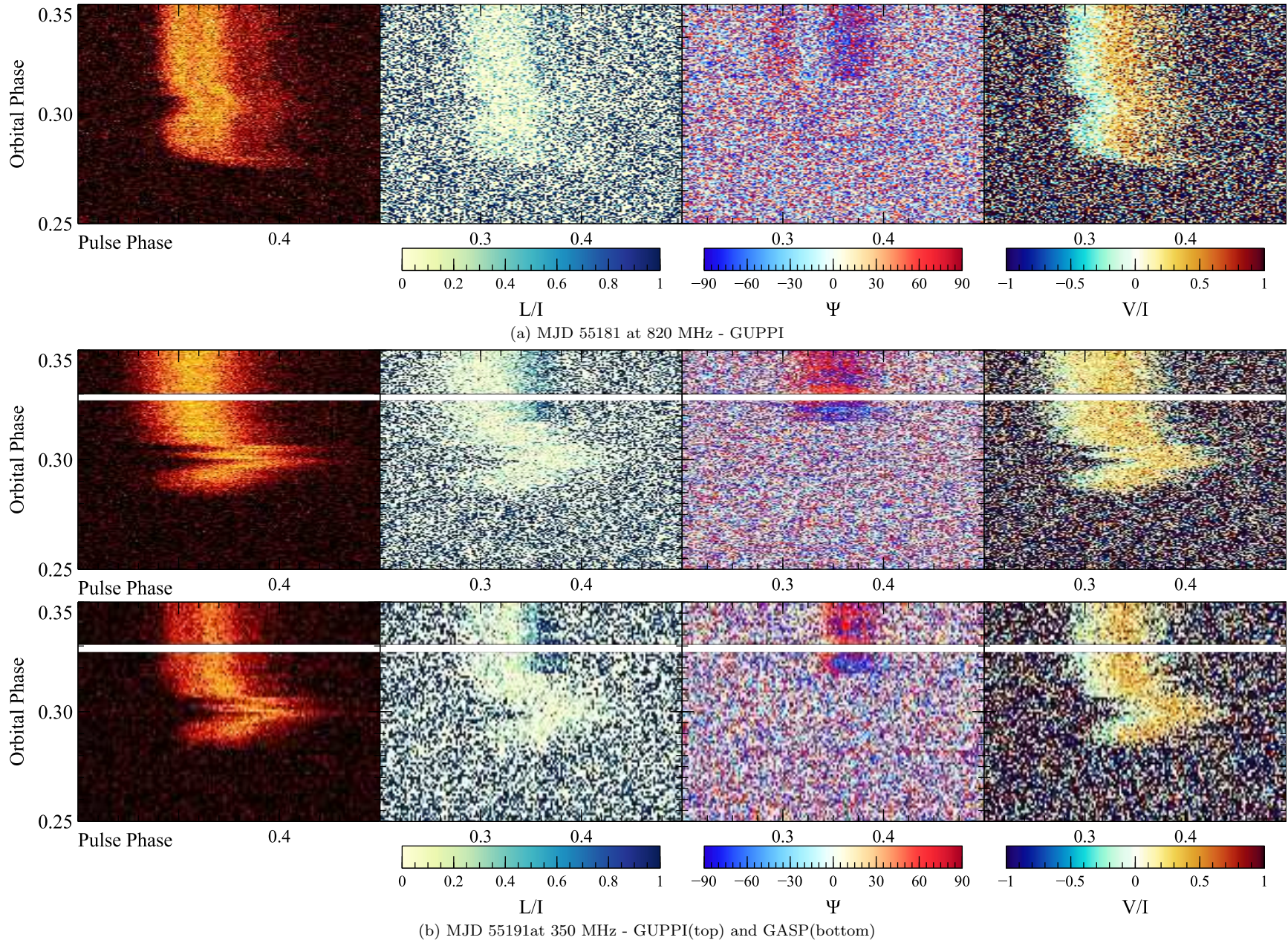}
        \caption{\label{fig:diff}Polarization parameters in the eclipse exit region. The leftmost plot shows the total intensity, I, for reference. Zero-weighted sub-integrations appear as white horizontal lines.}
	\end{figure}
\end{landscape}


\subsection{Polarization changes due to eclipsing material}
\label{subsec:pol_changes}

The original hope of this study was to look at polarization changes during the dispersive smear of the eclipse ingress and egress. As seen in \autoref{fig:higher_res}, both are fairly sharp and only three epochs cover this region of orbital phase. Observations taken at L-band on MJD 55226 had too low S/N for any variations to be visible. 
On MJDs 55181 and 55191, at \SI{820}{\MHz} and \SI{350}{\MHz} respectively, changes in the polarization profile were observed during the eclipse egress and the blip. No changes were discernible during ingress.

\autoref{fig:diff} shows unbinned close-ups of the eclipse egress and the blips in (left to right): total intensity (I), which has been included for reference, the fractional linear polarization (L/I), the polarization position angle ($\Psi$), and the fractional circular polarization (V/I). These are shown for MJDs 55181 and 55191. For MJD 55181, plots of GASP data have not been included as they show the same behaviour as the GUPPI plot.

In \autoref{fig:diff} the circular component of the polarization follows the total intensity and no deviations from I are apparent. However, we observe linear depolarization during eclipse egress and the blip. In both the GASP and GUPPI \SI{350}{\MHz} plot the peak in the linear polarization on the trailing edge of the profile is not present immediately after the eclipse; it then reappears at approximately 0.317 in orbital phase. Corresponding changes occur in the polarization position angle (PA) plot; a discernible PA profile, showing the orientation of the linear polarization, only reemerges from the noise at the same orbital phase. For the GUPPI MJD 55181 plot, while changes in the L/I plots are marginal or difficult to see, this same behaviour is clear in the plot of $\Psi$ (PA). 
From this we conclude that the clump or clumps causing the blip are linearly depolarizing the pulsar signal, perhaps due to a large or varying RM, but the circular polarization is not measurably affected. 

On MJD 55191, at \SI{350}{\MHz}, both GUPPI and GASP show a shift in the polarization position angle profile when the linear polarization reappears in the final part of the blip. As $\Psi$ is an orientation, with a range of 180\degree, this upward shift wraps the position angle which then appears in the negative end of the scale.  
Interestingly, we only observe a PA shift in the tail end of the blip when the DM has dropped much lower than its blip peak value. This PA shift is a clear indication of Faraday rotation due to the presence of a magnetic field with some component along the line-of-sight. Unfortunately, RFI was present in the sub-integrations between the shifted and non-shifted PA profiles. No similar shift can be seen in data taken on MJD 55181 at \SI{820}{\MHz}.

\begin{figure*}
    \centering
    \includegraphics[]{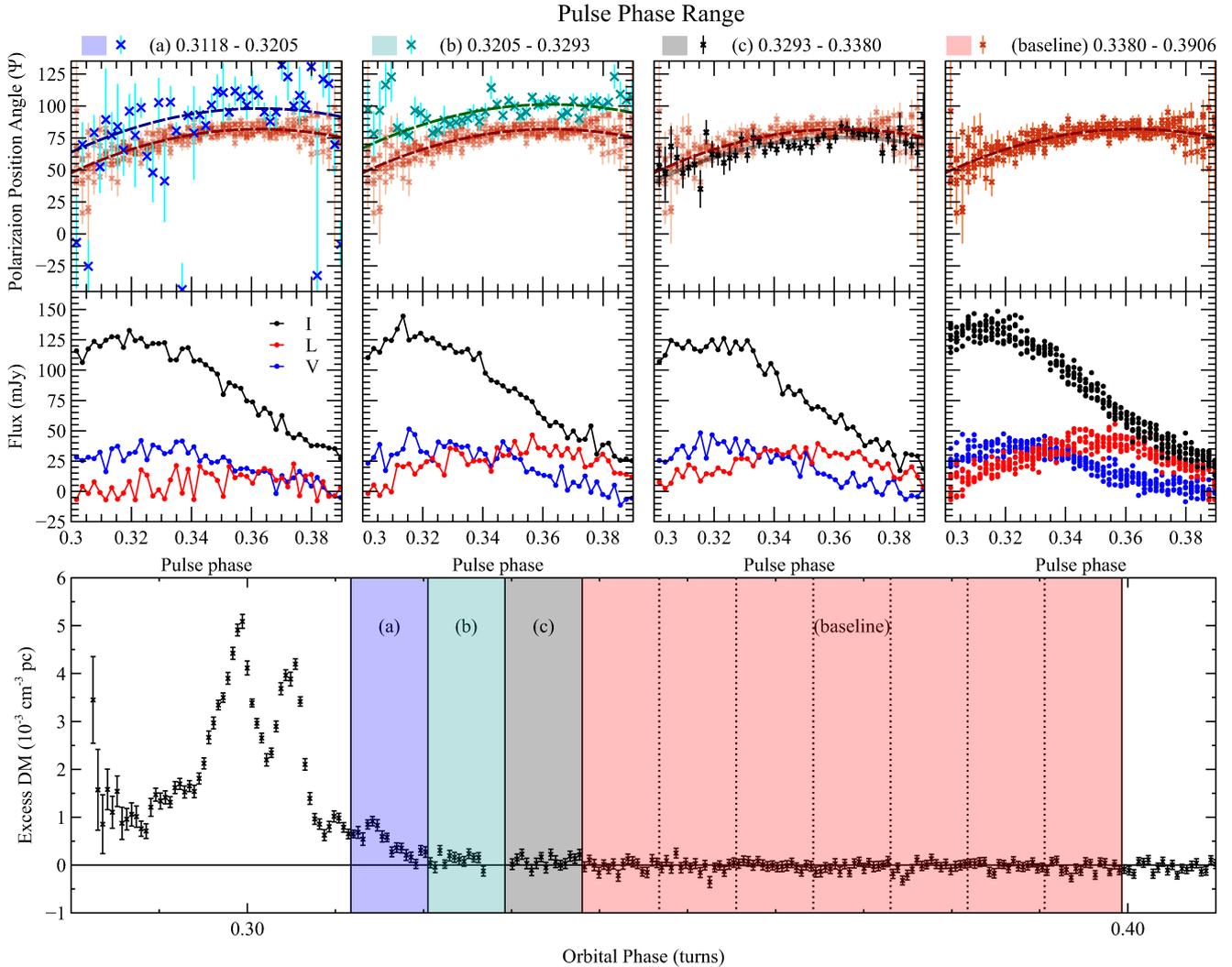}
    \caption{Measuring the PA shift on MJD 55191 at \SI{350}{\MHz} using GUPPI binned sub-integrations. (Top) PA and polarization profiles for each binned sub-integration; from left to right, (a), (b), (c), the 7 binned sub-integrations which form the baseline. The PA has been wrapped into the range \ang{-45} to \ang{135} for better visualization. The (baseline) PA profiles are included in all PA plots for reference. Fits shown in the PA plots were permitted to vary in y-intercept only and are not meant as true fits to the PA profiles. (Bottom) Excess DM formed from the same observation's higher resolution GUPPI residuals. This plot is included to show the location of each binned sub-integration with respect to the blip; it should be noted that excess DM values quoted in the text come from timing residuals of the binned sub-integrations, not the residuals shown in this plot. Shaded regions show the phase ranges of (a), (b), (c), and baseline; the baseline region is comprised of 7 binned sub-integrations whose limits are shown by dotted vertical lines. A horizontal line at 0 has been included to aid comparisons.}
    \label{fig:PAshift}
\end{figure*}

In order to measure the shift, the GUPPI 55191 observation shown in \autoref{fig:diff} was binned in time by a factor of 16 to increase the S/N, the same factor as was used to generate the TOAs for timing this observation. With this binning, two ``new" sub-integrations cover the PA shifted region. The intensities of the new sub-integrations were too low to measure RM using \prog{rmfit}, so the shift was measured by comparing PA profiles between the binned sub-integrations. These profiles are plotted in \autoref{fig:PAshift}.


\autoref{fig:PAshift} shows the PA and polarization profiles from the the binned sub-integrations with their uncertainties as output by PSRCHIVE. Also shown is excess DM data from the same observation, showing where the sub-integrations fall in orbital phase and with respect to the blip. A baseline PA profile was formed using PA profiles from 7 nearby sub-integrations, ranging from 0.3380 to 0.3906 in orbital phase, to minimize ionospheric RM variations between the shifted sub-integrations and the baseline. A quadratic function was fit to data from all 7 baseline sub-integrations over the pulse phase range shown. We assume there was no measurable change in the pulse profile shape and apply the same fit to (a) and (b) from \autoref{fig:PAshift}, covering 0.3118 - 0.3205 and 0.3205 -- 0.3292 in orbital phase respectively, allowing only the vertical offset to vary. (c) (0.3293 -- 0.3380 in orbital phase) shows a shape change from the baseline PA with a dip between approximately 0.33 and 0.36 in pulse phase. We do not know the cause of this shape change but due to this dip (c) is not included in the baseline sub-integrations.

We find fits to (a) and (b) are both statistically significantly offset from the baseline, but the points for (a) are far more scattered leading to a reduced chi squared statistic of 3.62 compared with $\chi^2_{\rm red} = 1.31$ for (b). Note that these are not ``true" fits as only the offset was permitted to vary but, to capture this difference in the scatter in some form, uncertainties were multiplied by $\sqrt{\chi^2_{\rm red}}$. 

In this way we find (a) is offset by \ang{16 \pm 4} and (b) by \ang{19 \pm 3} corresponding to rotation measures of \SI{0.38 \pm 0.10}{\rad\per\meter\squared} and \SI{0.46 \pm 0.07}{\rad\per\meter\squared} respectively. Combining these two measurements in a weighted mean gives an RM of \SI{0.44 \pm 0.06}{\rad\per\meter\squared}; this measurement is an excess RM in addition to the RM mentioned in \autoref{sec:pol_profiles}.

At \SI{820}{\MHz} this rotation measure would shift the PA profile by \ang{3.4 \pm 0.5}. If conditions on MJD 55181 produced a similar RM, given the low S/N and resulting scatter in the PA profile, this would explain our non-detection of a shift in the \SI{820}{\MHz} observation.

A simultaneous measurement of both dispersion and rotation measures can be used to calculate the magnetic field component along the line-of-sight. Computing an excess DM from the timing residuals, as described at the beginning of this section, for sub-integrations (a) and (b) gives \SI{0.49 \pm 0.02 E-3}{cm^{-3}pc} and \SI{0.16 \pm 0.07 E-3}{cm^{-3}pc} respectively. For comparison \SI{0.02 \pm 0.02 E-3}{cm^{-3}pc} is the mean excess DM magnitude for the baseline sub-integrations.

Sub-integration (a) has a poorly constrained RM measurement but a comparatively well constrained DM and vice versa for (b). As such we cannot identify variations in the magnetic field or any distance-dependence. However a magnetic-field measurement is still possible. Plus, with the caveat of low S/N and correspondingly large uncertainties, there are hints of interesting magnetic behaviour; the PA profile for (c) does deviate in shape from the baseline, and both (a) and (b) also suggest a changing profile shape. 

Combining the weighted mean RM with the DM from sub-integration (a), we measure a B-field of $\sim$\SI{1.11 \pm 0.16}{\milli\gauss}; using the DM from (b) gives $\sim$\SI{3.5 \pm 1.7}{\milli\gauss}. Both values are much larger than the Galactic magnetic field ($\approx \si{\micro\gauss}$~\citep{Jansson2012a,Jansson2012}). In addition there were no reported solar flares or ionospheric events on MJD 55191 which would imitate this effect. We believe this is the first successful detection of a non-zero magnetic field within the eclipsing material of a black widow or redback system. 

Previously \citet{Fruchter1990}, using the Faraday delay induced between left- and right-handed circular polarizations, measured a line-of-sight magnetic field for the original black widow pulsar, PSR B1957+20, of \SI[separate-uncertainty=true]{-1.5 \pm 4.5}{\gauss} and \SI[separate-uncertainty=true]{0.4 \pm 1.0}{\gauss} pre- and post-eclipse respectively. Effects from Faraday rotation are not seen in our circular polarization profiles. This is unsurprising as it is a smaller effect; following \citet[Equation~4]{Fruchter1990} we would expect a delay of \SI{0.14}{\nano\second} which is below our timing precision.

Polarization changes around pulsar eclipses have been observed before, for example in Ter5A where clumps of material remaining in the system and high eclipse variability were also observed~\citep{Anna2010}. \citet{Anna2010} also notes that as Ter5A enters eclipse, the linear polarization fades away before the circular polarization does so. For J2256--1024 we do not see any such phenomena but this may be due to the rapidity of the eclipse ingress. \citet{Anna2010} also notes a large amount of variability in the measured RM for Ter5A; it seems to be a good candidate for other magnetic field measurements.

Native time-resolution sub-integrations show shifted PA profiles start at \num{0.3175 \pm 0.0003} in orbital phase. The minimum distance between the companion and the ionized material, in which we measure the magnetic field, is the distance between the pulsar and the companion, at the time of the measurement, projected onto the plane of the sky. Assuming $i = \ang{90}$ this minimum distance is \SI{1.614 \pm 0.005}{\lightsecond} or \num{3.16}\,$R_{\rm L}^{\rm c}$, where $R_{\rm L}^{\rm c}$ is the effective Roche lobe radius of the companion.
Using the \citet{1302.1790} inclination angle of $\ang{68 \pm 11}$ gives a minimum distance of \SI{2.1 \pm 0.4}{\lightsecond} (\num{4.0}\,$R_{\rm L}^{\rm c}$).

An obvious candidate for the source of this magnetic field is the companion. As an estimate, we assume the companion has a dipolar magnetic field, a radius of $R_{\rm L}^{\rm c}$, and use the minimum distances to the measured $\sim$\SI{1.11}{\milli\gauss} field given in the previous paragraph. This implies the companion has a surface magnetic field of $\sim$\SI{35}{\milli\gauss} ($i = \ang{90}$) or $\sim$\SI{72}{\milli\gauss} ($i = \ang{68}$). However, requiring a pressure balance between the pulsar wind and the companion's magnetosphere (e.g. \citet{Harding1990,Wadiasingh2018}), at a companion surface located at $R_{\rm L}^{\rm c}$, gives $\sim$\SI{15}{\gauss} for the companion's surface magnetic field. Here we have assumed an isotropic pulsar wind with pressure $ \nicefrac{\dot E_{\rm rot}}{4\pi c (a-R_{\rm L}^{\rm c})^2}$ at the companion's closest surface to the pulsar, and a magnetic pressure from the companion's field of $\nicefrac{B^2}{8\pi}$.

Given that a) the pulsar wind is unlikely to be isotropic, b) it is unlikely \SI{100}{\percent} of the spin-down power is converted into wind,  c) there is likely a non-zero component of the B-field in the plane of the sky, and d) this calculation used the minimum distance between the ionized material and the companion, we believe the companion is still a reasonable source for the measured magnetic field. Combining the two calculations above -  a pressure balance at the companion surface and that the field drops to $\sim$\SI{1.11}{\milli\gauss} at \SI{1.614}{\lightsecond} ($i = \ang{90}$) / \SI{2.1}{\lightsecond} ($i = \ang{68}$) from the companion's center - to solve for the companion radius gives $0.14\, R_{\rm L}^{\rm c}$ ($i = \ang{90}$) / $0.18\, R_{\rm L}^{\rm c}$ ($i = \ang{68}$). We present these values as minimum radii for the companion, presuming it is the source of our measured magnetic field and its field is dipolar.

\section{Conclusions}
We find J2256--1024 to be a classic black widow pulsar with a low minimum mass companion of \SI{0.03}{\msun} in a tight orbit with a pulsar-companion separation \begin{math} \approx 7.6 \end{math} times the companion's effective Roche lobe radius. We present a tentative parallax measurement which yields a distance, \SI{2.0 \pm 0.6}{\kilo\parsec}, consistent with that inferred from the DM measurement using the YMW16 model - \SI{1.3 \pm 0.4}{\kilo\parsec}. 

The data span $\sim$3 years and observing epochs are unevenly distributed over that range - in particular there is a 341 day gap. As such, we were unable to fit a reliable proper motion and only give an upper limit. In addition only one spin frequency derivative and no orbital period derivatives were fit. These are natural targets for future study, particularly as orbital evolution and mass loss from a black widow system is expected. Multi-frequency observations would allow DM variations to be fitted, further improving a timing solution, and investigations into the frequency evolution of the polarization profile.

We see indications that the material ``blown" from the companion is clumpy, observing clumps on two epochs. In these clump events we observe linear depolarization of the polarization and, on one epoch, evidence of Faraday rotation due to the system's environment with an excess RM of \SI{0.44 \pm 0.06}{\rad\per\meter\squared}, leading to a line-of-sight magnetic field measurement of  $\sim$\SI{1.11 \pm 0.16}{\milli\gauss}. We believe this to be the first non-zero measurement of a magnetic field within eclipsing material in a black widow system and that the companion is a plausible source for the field.

Excess dispersion events have been observed in other black widow systems and redbacks. Investigations into their polarization properties seems a rich area for further study.

With regards to J2256--1024, observations at low frequencies around the eclipse region could provide insight into the frequency of such clumps and shed light on the nature of the measured magnetic field. There are few studies on pulsar wind and its interaction with matter in this regime so close to the pulsar as most focus on pulsar wind nebulae, a notable exception being \citet{AlessandroRidolfi2012}. J2256--1024 and other such systems could provide a useful constraint and insight into this process and the stripping of material from pulsar companions.

\section*{Acknowledgements}

The Green Bank Observatory is a facility of the National Science Foundation operated under cooperative agreement by Associated Universities, Inc. Data used in this analysis were taken under GBT projects AGBT09B\_031, AGBT09C\_072, AGBT10A\_060, AGBT10A\_082, AGBT10B\_018, AGBT11B\_070, and AGBT12A\_388.

The National Radio Astronomy Observatory is a facility of the National Science Foundation operated under cooperative agreement by Associated Universities, Inc. SMR is a CIFAR Senior Fellow. DRL, RSL, MAM, SMR, and KS are supported by the NSF Physics Frontiers Center award 1430284.

Pulsar research at UBC is supported by an NSERC Discovery Grant and by the Canadian Institute for Advanced Research.

JvL acknowledges funding from ERC Consolidator Grant 617199 (`ALERT') and from NWO Vici Grant  639.043.815 (`ARGO').



\bibliographystyle{mnras}
\bibliography{Mendeley,Fermi} 






\bsp	
\label{lastpage}
\end{document}

%% file: macros.tex
\let\sun\odot

\DeclareSIUnit{\msun}{\ensuremath{M_\sun}}
\DeclareSIUnit{\rsun}{\ensuremath{R_\sun}}
\DeclareSIUnit{\jansky}{Jy}
\DeclareSIUnit{\parsec}{pc}
\DeclareSIUnit{\au}{AU}
\DeclareSIUnit{\lightsecond}{lt \text{-} s} 
\DeclareSIUnit{\gauss}{G}
\DeclareSIUnit{\erg}{erg}

\DeclareSIUnit{\yearunit}{yr} 
\DeclareSIUnit{\hour}{hr}

\DeclareSIUnit{\masyr}{mas\per\yearunit} 
\DeclareSIUnit{\mas}{mas} 

\DeclareSIUnit{\rad}{rad}
\DeclareSIUnit{\dyne}{dyn}

\newcommand*{\ra}[2][]{{
		\def\SIUnitSymbolDegree{\textsuperscript{h}}%
		\def\SIUnitSymbolArcminute{\textsuperscript{m}}%
		\def\SIUnitSymbolArcsecond{\textsuperscript{s}}%
		\ang[#1]{#2}}%
}

\newcommand*{\dec}[2][]{{
		\def\SIUnitSymbolDegree{\textsuperscript{\degree}}%
		\def\SIUnitSymbolArcminute{\textsuperscript{m}}%
		\def\SIUnitSymbolArcsecond{\textsuperscript{s}}%
		\ang[#1]{#2}}%
}

\newcommand{\prog}[1]{\texttt{#1}}